\documentclass[prb,twocolumn,amsmath,amssymb,showpacs]{revtex4}
\usepackage{graphicx}

\def\beq{\begin{eqnarray}}
\def\eeq{\end{eqnarray}}
\def\beqa{\begin{eqnarray}}
\def\eeqa{\end{eqnarray}}

\begin{document}

\title{Spin exchange and superconductivity in a $t-J'-V$ model
for two-dimensional quarter-filled systems}
\author{Andr\'es Greco$^1$, Jaime Merino$^2$, Adriana Foussats$^3$ and Ross H.
McKenzie$^4$}
\affiliation{$^{1,3}$Facultad de Ciencias Exactas, Ingenier\'{\i}a y Agrimensura and Instituto de F\'{\i}sica Rosario (UNR-CONICET).  
Av. Pellegrini 250-2000. Rosario-Argentina. \\
$^2$Departamento de F\'isica Te\'orica de la Materia
Condensada, Universidad Aut\'onoma de Madrid, Madrid 28049, Spain.\\
$^4$ Department of Physics, University of Queensland, Brisbane 4072,
Australia.}
\date{\today}
\begin{abstract}
The effect of antiferromagnetic spin fluctuations on
two-dimensional quarter-filled systems is studied theoretically.
An effective $t-J'-V$ model on a square lattice which accounts
for checkerboard charge fluctuations and next-nearest-neighbors
antiferromagnetic spin fluctuations is considered.  From
calculations based on large-$N$ theory on this model 
it is found that the exchange interaction, $J'$,  
increases the attraction between electrons in the d$_{xy}$ channel only,
so that both charge and spin fluctuations work cooperatively to
produce d$_{xy}$ pairing.   
\end{abstract}

\maketitle
\section{Introduction}
\label{Intro} The BEDT-TTF (bis-ethyleneditiotetrathiafulvalene)
family of quarter-filled layered organic materials with the
$\theta$, $\alpha$ and $\beta''$ arrangements of the molecules
display a subtle competition between metallic, charge ordered
insulating and superconducting phases \cite{Ishiguro,Jaime1}. They are
examples of strongly correlated electron systems 
for which their electronic states are theoretically
described by a 2D extended Hubbard model at 3/4-filling of
electrons (1/4-filling of holes) for the HOMO of the BEDT-TTF
molecules\cite{Kino}. The nearest-neighbors intermolecular
Coulomb interaction, $V$, is a crucial ingredient as, at
one-quarter filling, the on site Coulomb repulsion, $U$, by itself
cannot describe charge ordering phenomena\cite{Seo}. The extended
Hubbard model at this filling has been previously studied
\cite{Jaime1} through large-$N$ and slave-boson approaches in the
$U$-infinite limit as well as with exact diagonalization on small
clusters \cite{Matteo} at finite-$U$. Several issues related to
charge ordering phenomena have been addressed. A transition from a
metal to a checkerboard charge ordered insulating state at a
finite $V=V_c$ has been found. Close to this charge ordered phase,
superconductivity in the $d_{xy}$ channel appears\cite{Jaime2}
induced by strong charge fluctuations. Dynamical \cite{Jaime4}
properties of the metallic phase in the presence of short range
charge fluctuations have been found to be anomalous in agreement
with experimental data\cite{Jaime3}.

Large-$N$ methods and slave bosons are useful for the study of the
effect of charge fluctuations on various electronic properties as
they can be included at $O(1/N)$, however, spin fluctuations are
typically neglected unless complicated $O(1/N^2)$ contributions
are considered. The effect of spin fluctuations on
superconductivity in quarter-filled systems has been
addressed through RPA (Random Phase Approximation) calculations
which consider antiferromagnetic instabilities induced by the
on-site Coulomb interaction, $U$ [\onlinecite{Ogata}], finding that $d_{xy}$
superconductivity still prevails. However, well in the charge
ordered insulating phase it is known that the spins order
antiferromagnetically due to the presence of a next-nearest
neighbor spin exchange interaction\cite{Jaime1}. This spin interaction results
from a 'ring' exchange process appearing at fourth order in $t$
and acts along the diagonals of the square lattice reading:  $J'=4
t^4/9 V^3$ in the $U \rightarrow \infty$, and $V >> t$ limits.
Exact diagonalization on 16-site clusters indicate that the
$(\pi/2,\pi/2)$ antiferromagnetic spin arrangement follows closely the
$(\pi,\pi)$ checkerboard arrangement of the charge \cite{Ohta} as 
$V/t$ is increased from the metal to the charge ordered phase.
These results suggest that remnants of the exchange interaction,
$J'$, generated in the insulating phase can survive in the
metallic phase where short range charge ordering is present. It is
then the purpose of the present work to analyze the influence of
this exchange coupling, $J'$, on the superconducting instabilities
previously found \cite{Jaime2} induced by charge fluctuations. As
this $J'$ acts along the diagonals of the lattice it is
conceivable that similarly to the $d_{x^2-y^2}$ superconductivity
appearing in the $t-J$ model close to half-filling induced by
antiferromagnetic spin fluctuations, the $J'$ appearing in the
quarter-filled $t-J'-V$ model can induce $d_{xy}$ pairing.

The paper is organized as follows. In Sect. \ref{model}
we introduce the quarter-filled $t-J'-V$ model and provide
a phase diagram obtained from the large-$N$ approach used here.
In Sect. \ref{supercond} we discuss the superconducting 
phase and study in detail the effect of $J'$ on the pairing symmetry.
Finally in Sect. \ref{conclu} we summarize our results and point
out their connection to the electronic properties of quarter-filled 
layered organic molecular crystals.

\section{The $t-J'-V$ model and its phase diagram}
\label{model}

\begin{figure}
\begin{center}
\setlength{\unitlength}{1cm}
\includegraphics[width=6cm,angle=0.]{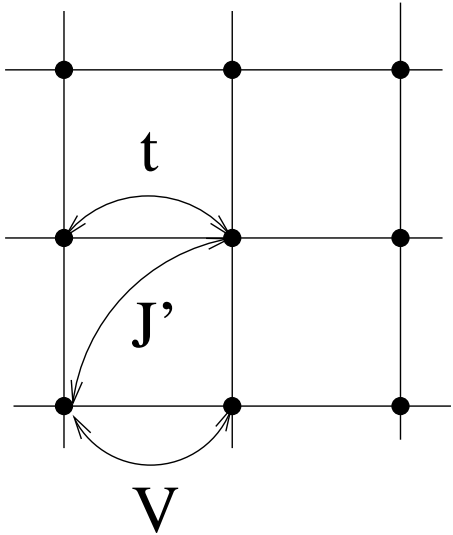}
\end{center}
\caption{A square lattice with a nearest-neighbors Coulomb repulsion $V$
and hopping amplitude $t$ and a next-nearest neighbors spin exchange 
coupling $J'$.}
\label{lattice}
\end{figure}

In order to explore the above possibility an effective
quarter-filled $t-J'-V$ model is introduced as a natural extension
of the extended Hubbard model previously studied (see Fig.
\ref{lattice} for a schematic sketch of the effective
interactions). The model reads
\begin{widetext}
\begin{eqnarray}
H = \sum_{<ij>,\sigma}\;(t_{ij}\; \tilde{c\dag}_{i\sigma}
\tilde{c}_{j \sigma} + h.c.) + \sum_{<ij>} J'_{ij} (\vec{S}_i
\cdot \vec{S}_j -\frac{1}{4}  n_i n_j ) + \sum_{<ij>} V_{ij} n_i
n_j,
\label{hamilt}
\end{eqnarray}
\end{widetext}

\noindent
where $t_{ij}$ and
$V_{ij}$ are the nearest-neighbors hopping and
Coulomb repulsion
parameters, respectively, between sites $i$ and $j$ on the square lattice.
$J'_{ij}$ is the antiferromagnetic
exchange interaction  between
second-neighbors sites.
$\tilde{c\dag}_{i \sigma}$ and $\tilde{c}_{i \sigma}$
are the fermionic creation and
destruction operators respectively under the constraint
that
double occupancies of lattice sites $i$ are excluded.
$\vec{S_i}$ and $n_i$ are the spin and
the fermionic density, respectively.

The model (\ref{hamilt}) is studied by using the large-$N$ approach
for Hubbard operators \cite{Adriana} recently  extended to the
case of finite $J$\cite{Adriana1} (see also
Ref.[\onlinecite{Jaime4}] for the $J=0$ case). In the appendix we
give details about this approach. The method has been thoroughly
tested on the $t-U-V$ model by comparing dynamical properties with
exact diagonalization calculations on small clusters
\cite{Jaime4}, finding good agreement for the behavior of charge
and spectral functions as well as spectral densities close to the
charge ordering transition. This agreement can be attributed to
the fact that the infinite-$U$ limit has been considered and
neglecting the nearest-neighbor exchange $J$ is justified since,
at finite $V$, the charge tries to sit in every other site of the
square lattice making $J$ ineffective.

Although the $t-J'-V$-model is only justified for values of $V$
sufficiently large, we have explored the full parameter range for
completeness. In the $t-J'-V$ model, the $J'$ is dynamically
generated when the charge is ordered within the checkerboard pattern
through a 'ring' exchange process. Hence, the $J'$ becomes
effective only when some sort of checkerboard charge ordering is
already present in the system (either short or long range charge order).
This means that the system should be charge ordered or
sufficiently close to the charge ordering transition for the
$t-J'-V$ model to be meaningful. The situation is different in the
$t-J$ model for the high-T$_c$ as, in this case, the $J$ is
generated through a super-exchange process at large on-site $U$.
\begin{figure}
\begin{center}
\setlength{\unitlength}{1cm}
\includegraphics[width=8cm,angle=0.]{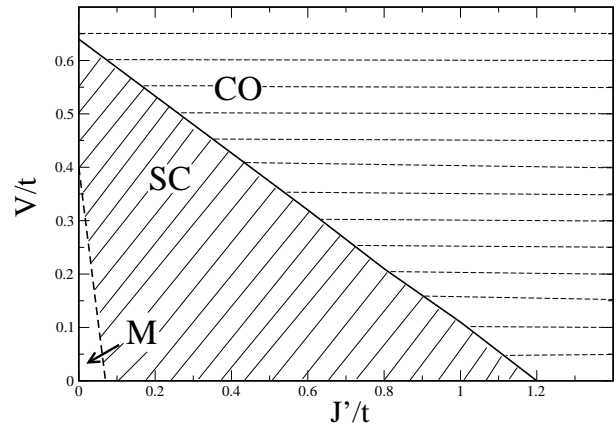}
\end{center}
\caption{Phase diagram obtained from the large-$N$ approach
in the $V-J'$ plane.  The solid line represents $V_c$ {\it i. e.} the critical
line signalling the onset of checkerboard charge ordering (CO).
Between the solid and dashed lines $d_{xy}$ superconductivity (SC) appears.
The region below the dashed line corresponds to the metallic (M) phase.
Superconductivity is found to be more robust closer to the CO
line and for larger $J'$.}
\label{fig2}
\end{figure}

A full phase diagram summarizing our results obtained from
large-$N$ theory on the $t-J'-V$ model is shown in Fig.
\ref{fig2}, where metallic (M), charge ordered (CO) and
superconducting (SC) phases occur. We start with a discussion of
the charge ordering transition. The critical value, $V_c$,
signalling the charge ordering (CO) transition of the metallic
phase as obtained from the divergence of the static charge
susceptibility (see appendix), is displayed as a solid line in
Fig. \ref{fig2}. For $J'=0$, the system charge orders at $V=V_c
\sim 0.65t$, as previously found \cite{Jaime1,Jaime4}. The value
of $V_c$ is found to decrease with increasing $J'$ which can be
easily understood from the following. By increasing $J'$,
checkerboard charge ordering is favored because two electrons are
antiferromagnetically attracted when they are sitting along the
diagonals of the square lattice so that a smaller $V$ is
effectively needed to induce CO. In a similar way, if we switch on
$J'$ for a fixed $V \lesssim V_c$, the CO phase is favored. The
dashed-dotted line in Fig. \ref{fig2} marks the onset for
superconductivity. Between solid and dashed lines the
superconducting effective coupling in the $d_{xy}$ channel (see
below) becomes negative indicative of superconducting pairing.
Superconductivity in the d$_{xy}$ channel is found to be more 
robust for larger $J'$ and closer to CO, as one would expect as in either 
way a stronger attraction is felt between quasiparticles in every other site 
of the lattice.

\begin{figure}
\begin{center}
\setlength{\unitlength}{1cm}
\includegraphics[width=8cm,angle=0.]{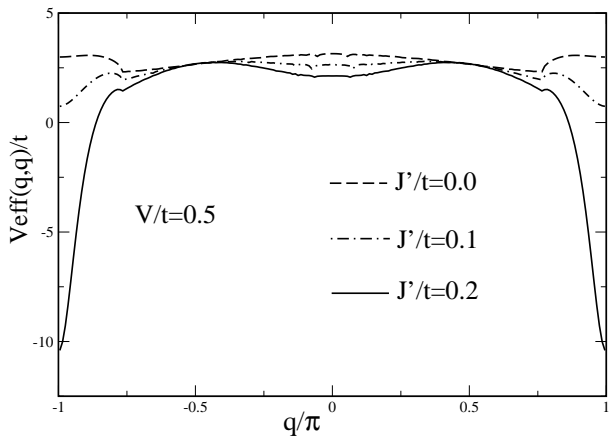}
\end{center}
\caption{Behavior of the effective potential between
quasiparticles, $V_{eff}(q,q)$, for $J'=0$, $J'=0.1t$ and
$J'=0.2t$, for fixed $V=0.5t$. For this parameters the system is
within the metallic phase close to the charge ordering transition
(see phase diagram in Fig. \ref{fig2}). As $J'$ is increased,
$V_{eff}(q,q)$ becomes more anisotropic and more attractive near
the momentum transfer ${\bf q} \sim(\pi,\pi)$.  Divergences
appearing at $(\pi,\pi)$ for increasing $J'$ favor
superconductivity in the $d_{xy}$ channel.  } \label{fig3}
\end{figure}

\section{Superconductivity in the $t-J'-V$ model}
\label{supercond}

In order to understand the model proposed we first discuss 
related large-$N$ studies performed in the well known models
such as the $t-J$ model relevant to the cuprates.
The case of a nearest-neighbors
exchange coupling, $J$, has been studied using a 
Slave-boson techniques have been used in combination 
with a $1/N$ expansion to study superconducting instabilities 
on $t-J$ model \cite{Grilli} and Hubbard model \cite{Kotliar}.  
Both the present approach (as described in the appendix) and the
slave-boson one, to $O(1)$, lead to fermions renormalized by the
presence of the Coulomb interaction. Indeed, superconductivity in the
model can only appear at $O(1/N)$. By adding an extra $J'$ term to the
bare extended Hubbard hamiltonian ($t-J'-V$-model) we can treat both 
$J'$ and $V$-terms at the same level of approximation, {\it i.
e.} through $O(1/N)$ and analyze possible superconducting
instabilities. A closely related work by Vojta \cite{Vojta}  
analyzes recently coexistence of superconductivity and checkerboard 
charge ordering within a $t-J-V$ model close to half-filling 
relevant to STM experiments on the cuprates\cite{Davis}.

Superconductivity is then investigated, within our model, by calculating
the effective interaction, $V_{eff}(\bf q)$, through $O(1/N)$ (see
appendix), between fermions for finite $J'$ at one-quarter
filling. The potential, $V_{eff}(\bf q)$ is plotted along the
${\bf q}=(q,q)$ direction in Fig. \ref{fig3} as a function of $q$
for different $J'$: $J'=0$, $J'=0.1t$ and $J'=0.2t$, and fixed
$V=0.5t$. For this set of parameters the system is always metallic
but close to the charge ordering transition as can be seen from
Fig. \ref{fig2}. For $J'=0$, $V_{eff}(\bf q)$ is repulsive and
anisotropic as found previously (see Fig. 13 of
Ref.[\onlinecite{Jaime4}]). As $J'$ is increased, $V_{eff}(q,q)$
becomes more anisotropic and more attractive near the momentum
transfer ${\bf q} \sim(\pi,\pi)$.  This behavior favors
superconductivity in the $d_{xy}$ channel as the $J'$ attracts the
charge tending to form the checkerboard pattern along the
diagonals of the square lattice.

There are two kinds of interactions contributing to $V_{eff}({\bf
q})$. For $J'=0$, the effective interaction close to CO is mainly
dominated by charge fluctuations associated with the collective
excitations near $(\pi,\pi)$ \cite{Jaime4}. Therefore, this kind
of pairing interaction, like the phonon mechanism in simple
metals, is mainly retarded and occurs in momentum space. For
$V=0$, the main effective interaction is of magnetic origin,
unretarded and short range in real space. For $J'$ and $V$
different from zero both kind of interactions contribute cooperatively 
to the binding energy of the Cooper pairs.

We use this effective potential to compute the effective couplings
in the different pairing channels or irreducible representations
of the order parameter, $i$ $(i=(d_{x^2-y^2}, d_{xy}, p))$. In
this way we can project out the interaction with a certain
symmetry. The critical temperatures, T$_c$, can then be estimated
in weak coupling from: T$_{ci}= 1.13 \omega_0
\exp(-{1/|\lambda_i|})$, where $\omega_0$ is a suitable cutoff
frequency and $\lambda_i$ are the effective couplings with
different symmetries. These are defined as \cite{Jaime4}:

\begin{widetext}
\begin{eqnarray}
\lambda_i=\frac{1}{(2 \pi)^2}
\frac{\int (d {\bf k} /|v_{\bf k}|) \int (d {\bf k'}/|v_{\bf k'}|)
g_i({\bf k'})
V_{eff}({\bf k'-k}) g_i({\bf k})}{
\int (d {\bf k}/|v_{\bf k}|)  g_i({\bf k})^2 }
\end{eqnarray}
\end{widetext}

\noindent
where the functions $g_i({\bf k})$, encode the different pairing
symmetries, and $v_{\bf k}$ are the quasiparticle velocities at
the Fermi surface. The integrations are restricted to the Fermi
surface. $\lambda_i$ measures the strength of the interaction
between electrons at the Fermi surface in a given symmetry channel
$i$. If $\lambda_i > 0$, electrons are repelled. Hence,
superconductivity is only possible when $\lambda_i <0$.

\begin{figure}
\begin{center}
\setlength{\unitlength}{1cm}
\includegraphics[width=8cm,angle=0.]{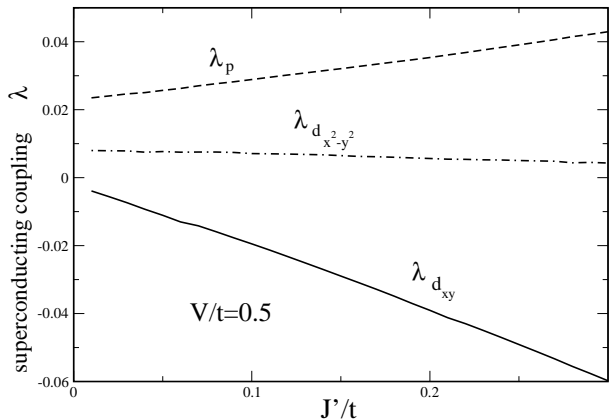}
\end{center}
\caption{
The dimensionless superconducting
coupling $\lambda$ in the $p$ ($\lambda_p$)(dashed line), $d_{x^2-y^2}$
($\lambda_{d_{x^2-y^2}}$) (dashed-dotted line) and
$d_{xy}$ ($\lambda_{d_{xy}}$)
(solid line) channels as a function of
$J'$ for a given value of $V=0.5 t$ in the metallic
phase close to the CO instability.
The second-neighbors antiferromagnetic exchange, $J'$, 
favors superconductivity in the $d_{xy}$ channel only.
}
\label{fig4}
\end{figure}

The coupling strength $\lambda$ for $d_{xy}$ superconductivity
has been found \cite{Jaime2,Jaime4}
to be very small (see Fig. 15 of Ref.[\onlinecite{Jaime4}]) previously.
For these small couplings, the corresponding superconducting critical
temperature T$_c$
is predicted to be very low. Other indications for pairing come from
exact diagonalization calculations of the binding energy
of two holes which becomes negative near the charge ordering transition
\cite{Jaime4}.
In Fig. \ref{fig4} we present results for
the dimensionless superconducting
coupling $\lambda$ in the $p$ ($\lambda_p$), $d_{x^2-y^2}$
($\lambda_{d_{x^2-y^2}}$) and $d_{xy}$ ($\lambda_{d_{xy}}$) channels
as a function of $J'$ for a given value of
$V=0.5t$ close to the CO instability.

Clearly, superconductivity becomes more favorable in the $d_{xy}$
channel only as $J'$ is increased. The influence of $J'$ on the
$d_{x^2-y^2}$-symmetry (dashed-dotted line in Fig. \ref{fig4}) is
rather weak. When $J'$ increases, superconducting couplings are
more repulsive in the $p$-channel and even more in the
$s$-symmetry (not shown) channel.

Note also the large difference between the $\lambda$ values for
$J'=0$ \cite{Jaime4} and $J'=0.3t$. For $J'=0.3t$,
$\lambda$ is one order of magnitude larger than for $J'=0$ at $V=0.5t$.
This behavior, which only occurs in the $d_{xy}$ channel, shows the
strong influence of the second neighbors effective antiferromagnetic exchange
coupling $J'$ on $d_{xy}$ superconductivity.

Although we find a substantial enhancement of pairing in the
$d_{xy}$ channel with $J'$, the associated T$_c$ is yet very low from simple
estimates: T$_c \sim  e^{-1/|\lambda_i|} \sim 6\cdot10^{-8} t$, which is
tiny even taking the most favorable case of $J'=0.3t$ (see Fig. \ref{fig4})
for pairing, for which, $\lambda_{d_{xy}} \sim -0.06$. Small T$_c$
values have been also found by Motrunich and Lee \cite{Motrunich} in
the context of Na$_x$CoO$_2$, although their couplings are typically 
larger than ours because,
in the case of Na$_x$CoO$_2$, several sections of the Fermi surface  
are connected by the charge ordering wavevector for  
$x=1/3$.  It is worth noting that in our approach we have 
not considered the renormalization of the quasiparticles, which enhances the 
effective mass, that occurs close to the charge ordering transition due to $V$.
This could be taken into account by including self-energy effects
in the calculation. One would then have to consider that the
hopping amplitudes are effectively renormalized by the
quasiparticle weight, $Z$, which decreases near the
charge instability \cite{Jaime4}. This would transform the bare
hopping, $t$, to $t_{eff}=Zt$. Hence, the ratios
$J/t_{eff}$ and $V/t_{eff}$ would be enhanced,
effectively increasing $|\lambda_{d_{xy}}|$ which would, in turn,
significantly increase the estimated T$_c$.

\section{Conclusions}
\label{conclu}

In conclusion, we have found that superconductivity with $d_{xy}$
symmetry induced by charge fluctuations is strengthened by
antiferromagnetic spin fluctuations induced by the exchange
coupling, $J'$, in a $t-J'-V$ model for quarter-filled systems.
This can be intuitively understood by analogy with the more
standard case of $d_{x^2-y^2}$ superconductivity induced by the
superexchange coupling $J$ in the $t-J$ model if one realizes that
the $J'$ acts in directions rotated by $45^0$ with respect to $J$.
Both charge and spin fluctuations are then found to work together
cooperatively to produce $d_{xy}$ superconductivity in the
quarter-filled $t-J'-V$ model. 

The way superconductivity behaves in the $t-J'-V$ model
proposed here, could be viewed as a two step process in which the charge
fluctuations are responsible for the onset of SC in the first
place and subsequently the dynamically generated spin exchange
coupling $J'$ would strengthen the binding between electrons
forming the Cooper pairs. 

Critical temperatures are found to be too small compared to
experimental values (which are of the order of a few Kelvin). 
The smallness of T$_c$ is related to the small Fermi surface
associated with the one-quarter filling of the system. Due to this
fact, there are no two points at the Fermi surface connected by the
$(\pi,\pi)$ CO wavevector which makes the interaction less effective in
producing Cooper pairs than in nearly antiferromagnetic metals close to
half-filling. A similar situation arises in the $t-J$ model \cite{Grilli}
at one-quarter filling in which superconductivity in the d$_{x^2-y^2}$ channel
is found although with rather small attractive effective couplings.
Only at dopings close to half-filling (for doping levels of
at most 0.15-0.2) the couplings are found to be substantial \cite{Andres}.
This is because, in this case, larger Fermi surface sections are effectively connected by the AF $(\pi,\pi)$ wavevector.

An important finding that derives from our work is that
including $J'$ is crucial in order to enhance
T$_c$ sufficiently. If only the $V$ is taken into
account, the values of T$_c$\cite{Jaime4} obtained would
be astronomically small as previously noted\cite{Jaime4}.  
A more sophisticated theory including the 
renormalization of the quasiparticles could 
enhance the estimates for T$_c$ even further.  Measured effective
masses in quarter-filled layered organics correspond to values of about:
$m^*/m=1-2$, leading to $J'/t$ larger by a factor of 2. 
Considering the dependence of $\lambda_{d_{xy}}$ with $J'/t$
shown in Fig. \ref{fig4}, this many-body effect can lead 
to large enhancements in T$_c$ as T$_c$ depends exponentially 
with $\lambda_{d_{xy}}$. 
Hence, T$_c$'s of a fraction of a Kelvin can be obtained in the
most favorable case, for $V=0.5t$ and $J/t=0.3t$, considering $t\sim 0.1$
eV \cite{Jaime1}.

It has been recently suggested by Coldea {\it et. al.} 
Ref. [\onlinecite{Coldea}] that certain experimental observations are 
consistent with the charge mediated superconductivity scenario.  
For instance, the unit cell volume of 
$\beta''$-(BEDT-TTF)$_4$[(H$_3$O)M(C$_2$O$_4$)$_3$]$\cdot$ Y
[\onlinecite{Coldea}] and $\alpha$-(BEDT-TTF)$_2$MHg(SCN)$_4$
salts [\onlinecite{Drichko}], is found to increase from metallic
to insulating salts by changing M and Y. Superconducting salts such as
$\alpha$-(BEDT-TTF)$_2$NH$_4$Hg(SCN)$_4$ and the recently analyzed
$\beta''$-(BEDT-TTF)$_4$[(H$_3$O)Ga(C$_2$O$_4$)$_3$].C$_6$H$_5$NO$_2$
are found to have unit cell volumes right between their respective
metallic and insulating salts.  Increasing the unit-cell volume 
is translated to an increase in $V/t$ as well as in $J'/t$ within our 
model which drives the system closer to the charge ordered state. 
In the critical region between the metal and charge ordered phase, 
$V \lesssim V_c$, superconductivity is predicted to appear \cite{Jaime2}.  

However, a definitive test for unconventional $d$-wave pairing
in quasi-two-dimensional quarter-filled organics is yet missing.  
Possible experimental probes could come from measurements of the
Knight shift and NMR relaxation rate in the metallic phase close 
to the CO transition. In contrast to the $\kappa$-(BEDT-TTF)$_2$X 
superconductors, in which large enhancements of the Korringa 
ratio are found due to their closeness to a Mott phase, 
in the quarter-filled systems studied here there should be no 
enhancement of the Korringa ratio. Finally, the dependence of 
T$_c$ on impurities and disorder can also be used to distinguish
$d$-wave superconductivity from conventional $s$-wave pairing
as recently pointed out\cite{Powell}.  
 
\acknowledgments

J. M. acknowledges financial support from the Ministerio de
Ciencia y Tecnolog\'ia through the Ram\'on y Cajal program and EU
under contract MERG-CT-2004-506177. We acknowledge helpful discussions
with L. Manuel, J. Riera, A. Trumper and M. Vojta. R. H. M. acknowledges 
financial support from the Australian Research Council.

\appendix
\section{Appendix: Large-$N$ approach for $t-J'-V$ model}


In this appendix, we will give a summary of the path integral
large-N approach for Hubbard operators
\cite{Adriana,Jaime4,Adriana1} used in the present paper.
More detailed version of the formalism including the 
exchange coupling can be found in \cite{Adriana1}.

First we introduce Hubbard operators which are related with the
usual fermionic operators by

\begin{widetext}
\begin{eqnarray}
X_{i}^{0 \sigma} = (1-{c}_{i \bar{\sigma}}^{\dagger}
c_{i \bar{\sigma}}) c_{i \sigma},
 \ \ \ X_{i}^{\sigma 0}= (X_{i}^{0 \sigma})^{\dagger} ,
\ \  \ X_{i}^{\sigma \sigma'}={c_{i\sigma}^{\dagger}}
{c}_{i \sigma'}.
\end{eqnarray}
\end{widetext}

The five Hubbard ${X_i}$-operators
 $X_i^{\sigma \sigma'}$ and $X_i^{0 0}$ are boson-like
and the four Hubbard $X$-operators $X_i^{\sigma 0}$ and
$X_i^{0 \sigma}$ are fermion-like. The names fermion-like and
boson-like come from the fact that Hubbard operators do not verify
the usual fermionic and bosonic commutation
relations\cite{Hubbard}.

From the above relations we note that
$X^{\sigma 0}_i=\tilde{c\dag}_{i \sigma}$
and $X^{\uparrow \downarrow}_i=S^{+}_i$.

The Hubbard operators satisfy:

a)the completeness condition

\begin{eqnarray}
X_{i}^{0 0} + \sum_{\sigma} X_{i}^{\sigma \sigma} = 1 ,
\end{eqnarray}

\noindent
which is equivalent to imposing
that "double occupancy" at each site is forbidden.

b)the commutation rules

\begin{eqnarray}
[X_i^{\alpha \beta},X_j^{\gamma \delta}]_{\pm}= \delta_{ij}
(\delta^{\beta \gamma} X_i^{\alpha \delta} \pm \delta^{\alpha
\delta} X_i^{\gamma \beta})
\end{eqnarray}

\noindent
where the $+$ sign must be used when both operators are fermion-like,
otherwise it corresponds the $-$ sign.

On the basis of Hubbard $X$-operators,
the $t-J'-V$ Hamiltonian (1) is of the form:

\begin{widetext}
\begin{eqnarray}
H(X) = \sum_{<ij>,\sigma}\;(t_{ij}\; X_{i}^{\sigma 0}
X_{j}^{0
\sigma} + h.c.)
+ \frac{1}{2} \sum_{<ij>;\sigma} J'_{ij}
(X_{i}^{\sigma
{\bar{\sigma}}} X_{j}^{{\bar{\sigma}} \sigma} -
X_{i}^{\sigma
\sigma} X_{j}^{\bar{\sigma} \bar{\sigma}})
+ \sum_{<ij>;\sigma \sigma'} V_{ij} X_{i}^{\sigma \sigma}
X_{j}^{\sigma' \sigma'}
-\mu\sum_{i,\sigma}\;X_i^{\sigma \sigma}
\end{eqnarray}
\end{widetext}

\begin{widetext}
Our starting point is the path integral partition function
$Z$ written in the
Euclidean form
\end{widetext}

\begin{widetext}
\begin{eqnarray}
Z=\int {\cal D}X_{i}^{\alpha \beta}\;
\delta[X_{i}^{0 0} + \sum_{\sigma} X_{i}^{\sigma \sigma}-1]\;
\delta[X_{i}^{\sigma \sigma'} - \frac{X_{i}^{\sigma 0}
X_{i}^{0 \sigma'}}{X_{i}^{0 0}}]
\times({\rm sdet} M_{AB})_{i}^{\frac{1}{2}}
\exp\;(- \int d\tau\;L_E(X, \dot{X}))
\end{eqnarray}
\end{widetext}

The Euclidean Lagrangian $L_E(X,\dot{X})$ in (A5) is

\begin{eqnarray}
L_E(X, \dot{X}) =  \frac{1}{2}
\sum_{i, \sigma}\frac{({\dot{X_{i}}}^{0 \sigma}\;X_{i}^{\sigma 0}
+ {\dot{X_{i}}}^{\sigma 0}\;
X_{i}^{0 \sigma})} {X_{i}^{0 0}} + H(X)
\end{eqnarray}

In this path integral we associate Grassmann and usual
bosonic variables with Fermi-like and boson-like
$X$-operators, respectively.

It is worth noting at this point that the path integral
representation of the partition function (A5), looks
different to that usually found in other solid state problems. The
measure of the integral contains additional constraints  as well
as a determinant, $(sdet M_{AB})_{i}^{\frac{1}{2}}$.  Also the
kinetic term of the Lagrangian (A6) is
non-polynomial. The determinant reads

\begin{eqnarray}
(sdet M_{AB})_{i}^{\frac{1}{2}}=1/\frac{1}{(-X^{00})^2},
\end{eqnarray}

and is formed by all the constraints of the theory.

We now discuss the main steps needed to introduce a large-$N$
expansion of the partition function (A5). First, we
integrate over the boson variables $X^{\sigma \sigma'}$ using the
second $\delta$-function in (A5).  We extend the spin
index $\sigma=\pm$, to a new index $p$ running from $1$ to $N$. In
order to get a finite theory in the $N\rightarrow \infty$ limit,
we re-scale $t_{ij}$ to $t_{ij}/N$,
$V_{i j}$ to
$V_{i j}/N$ and
$J'_{i j}$ to
$J'_{i j}/N$.
The
completeness condition is enforced by exponentiating $X_{i}^{0 0}
+ \sum_{p} X_{i}^{p p} = N/2 $, with the help of Lagrangian
multipliers $\lambda_i$. We write the boson fields in terms of
static mean-field values, $(r_0, \lambda_0)$ and dynamic
fluctuations

\begin{eqnarray}
X_{i}^{0 0} &=& N r_{0}(1 + \delta R_{i}) \nonumber \\
\lambda_{i} &=&\lambda_{0}+ \delta{\lambda_{i}},
\end{eqnarray}

and, we make the following change of variables

\begin{eqnarray}
f^{+}_{i p} &=& \frac{1}{\sqrt{N r_{0}}}X_{i}^{p 0} \nonumber \\
f_{i p} &=& \frac{1}{\sqrt{N r_{0}}}\;X_{i}^{0 p},
\end{eqnarray}

where $f^{+}_{i p}$ and $f_{i p}$ are Grassmann variables.

The exchange interactions can be decoupled in terms of the
bond variable $\Delta_{ij}$ through a Hubbard-Stratonovich transformation,
where  $\Delta_{ij}$ is
the field associated with the quantity
$\sum_{p} \frac{f^{+}_{j p} f_{i p}}{ \sqrt{(1 + \delta R_{i})
(1 + \delta R_{j})}}$. We write the $\Delta_{ij}$ fields in term of
static mean field values and dynamics fluctuations
$\Delta_i^{\eta}=\Delta(1+r_i^\eta+iA_i^\eta)$,  where $\eta$ can take
two values associated with the bond directions $\eta_1=(1,1)$ and
$\eta_2=(-1,1)$ in real space.

\begin{figure}
\begin{center}
\setlength{\unitlength}{1cm}
\includegraphics[width=8cm,angle=0]{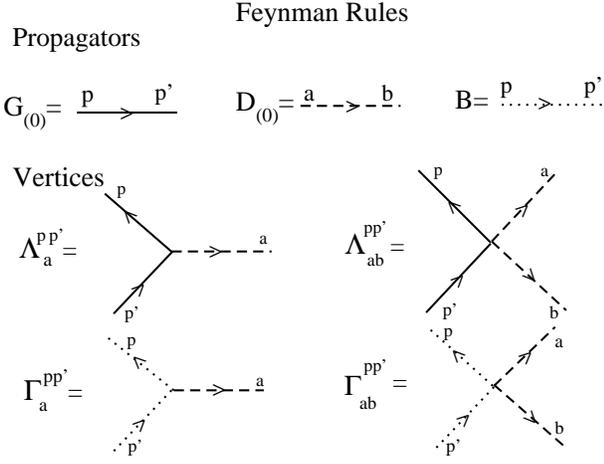}
\end{center}
\caption{Summary of the Feynman rules. Solid line represents the
propagator $G_{(0)}$ for the correlated fermion $X^{0\sigma}$.
Dashed line
represents the $6 \times 6$ boson propagator
$D_{(0)}$
for the $6$-component field $\delta X^a$. The component
$(1,1)$ of this propagator is directly associated
with the $X^{00}$ charge operator. Doted line is the propagator
$B$ for the boson ghost field ${\cal Z}_p$.
$\Lambda^{pp'}_a$ and $\Lambda^{pp'}_{ab}$ represent
the interaction between two fermions $f_p$ and one and two bosons
$\delta X^a$ respectively.
$\Gamma^{pp'}_a$ and $\Gamma^{pp'}_{ab}$ represent
the interaction between two ghost fields ${\cal Z}_p$ and one and two bosons
$\delta X^a$ respectively.
}
\label{fig5}
\end{figure}

Introducing the above change of variables
into Eq. (A5) and, after expanding the
denominators $1/(1+\delta R)$, we arrive at the following
effective Lagrangian:

\begin{widetext}
\begin{eqnarray}
&L_{eff}&=-\frac{1}{2}\sum_{i,p}\left(\dot{f_{i p}}f^{+}_{i p}
+ \dot{f^{+}_{i p}}f_{i p}\right) (1 - \delta R_{i} + \delta
R_{i}^{2})
+\sum_{<ij>,p}\;(t_{ij} r_{0} f^{+}_{i
p}f_{j p}+h.c.) - \mu \;\sum_{i,p}\;f^{+}_{i p}f_{i p} (1 - \delta R_{i}
+ \delta R_{i}^{2}) \nonumber \\ & + &
N\;r_{0}\;\sum_{i}\delta{\lambda_{i}}\;\delta R_{i} +\sum_{i,p}
f^{+}_{i p}f_{i p}(1 - \delta R_{i}) \;
\delta{\lambda_{i}} + \frac{2 N}{J}
\Delta^{2}\sum_{i\eta}\left[({r_{i}^{\eta}})^{2}+
({A_{i}^{\eta}})^{2}\right]
\nonumber
\\
& -&\Delta \sum_{<ij>,p, p'} (f^{+}_{i p}f_{j p'}+f^{+}_{j p'}f_{i
p})[1 -\frac{1}{2} (\delta R_{i} + \delta
R_{j})+\frac{1}{4}\delta R_{i}\delta R_{j}+\frac{3}{8}(\delta
R_{i}^{2} + \delta R_{j}^{2})] \nonumber
\\ & -&\Delta \sum_{<ij>,p, p'} (f^{+}_{i p}f_{j p'}+f^{+}_{j
p'}f_{i p})({r_{i}^{\eta}}+i {A_{i}^{\eta}})[1 -\frac{1}{2}
(\delta R_{i} + \delta R_{j})] \nonumber
\\
& + &  Nr_{0}^{2}   \sum_{<ij>}  (V_{ij}-  \frac{1}{2}J_{ij})
\delta R_{i}\delta R_{j}
 -\sum_{i p}\; {\bf {\cal Z}}_{i
p}^{\dag}\left(1- \delta R_i+ \delta R^2_i\right) {\bf {\cal
Z}}_{i p},
\end{eqnarray}
\end{widetext}

\noindent
where we have changed $\mu$ to $\mu-\lambda_0$ and dropped
constant and linear terms in the fields.

The last term of (A10) results from the path integral representation
of the determinant which uses $N$-component boson ghost field
${\cal Z}_p$
\cite{Adriana,Jaime4,Adriana1}.

Looking at the effective Lagrangian (A10), the Feynman rules can be
obtained as usual. The bilinear parts give rise to the propagators
and the remaining pieces are represented by vertices. Besides, we
assume the equation (A10) written in the momentum space once the
Fourier transformation was performed.

To leading order of $1/N$, we associate with the $N$-component
fermion field $f_{p}$, connecting two generic components p and p',
the propagator

\begin{eqnarray}
G_{(0)pp'}({\bf k}, \nu_{n}) = - \frac{\delta_{pp'}}{i\nu_{n} -
(E_{\bf k} - \mu )}
\end{eqnarray}

\noindent which is $O(1)$ and where $E_{\bf k}=
-2tr_{0}(cosk_x+cosk_y)-2\Delta cosk_xcosk_y$, is the electronic
dispersion to leading order.

The quantities ${\bf k}$ and $\nu_{n}$ are the momentum and the
fermionic  Matsubara frequency of the fermionic field,
respectively.

The mean field values $r_0$ and $\Delta$ must be determined
minimizing the leading order theory. From (A8) and the
completeness condition,  $r_0$ is equal to $\delta/2$, where
$\delta$ is the hole doping away from half-filling.

On the other hand, minimizing with respect to $\Delta$ we obtain
$\Delta= \frac{J'}{2} \frac{1}{N_s} \sum_{\bf k}
cos{k_x}\;\cos{k_{y}} n_F(E_{\bf k}-\mu)$, where $n_F$ is the
Fermi function and $N_s$ is the number of sites in the Brillouin
zone (BZ).

For a given doping, $\delta$,
the chemical potential $\mu$ and $\Delta$
must be determined self-consistently from
$(1-\delta)=\frac{2}{N_s} \sum_{k} n_F(E_k-\mu)$

\begin{figure}
\begin{center}
\setlength{\unitlength}{4cm}
\includegraphics[width=8cm,angle=0]{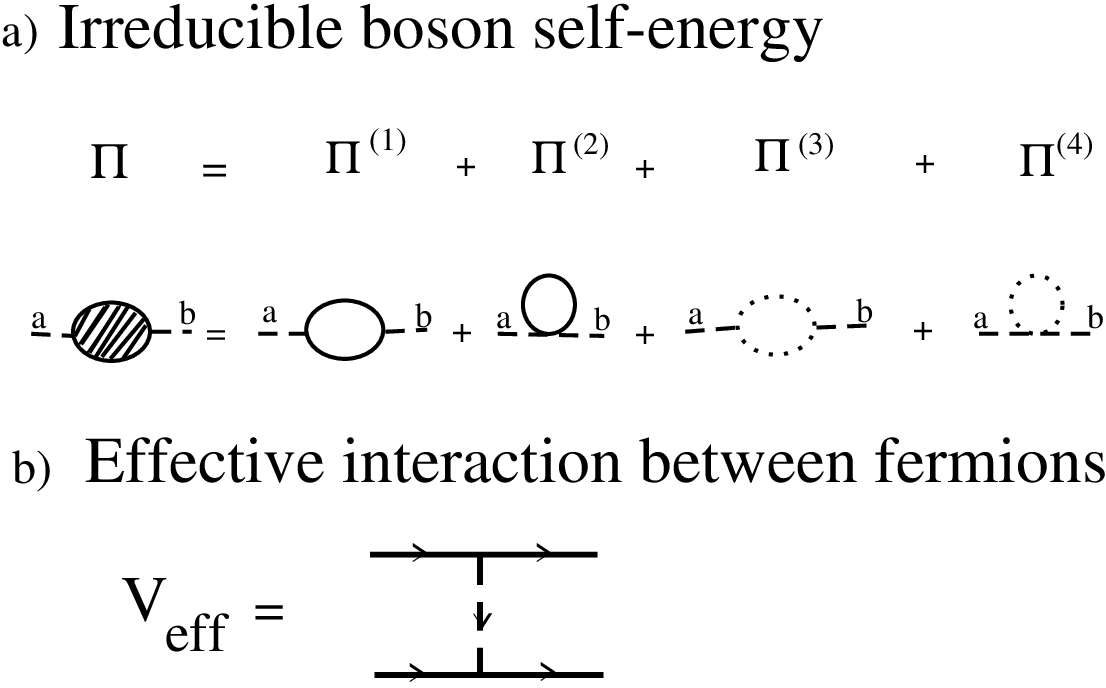}
\end{center}
\caption{a) The four different contributions
$\Pi^{(i)}_{ab}$ ($i=1,2,3,4$) to
the irreducible boson self-energy $\Pi_{ab}$.
b) Effective interaction between fermions. Only two three-legs vertices
contribute.
}
\label{fig6}
\end{figure}

We associate with the six component $\delta X^{a} = (\delta
R\;,\;\delta{\lambda},\; r^{1},\;r^{2},\; A^{1},\; A^{2})$
the inverse of the
propagator (which is $O(1/N)$),
connecting two generic components a and b,

\begin{widetext}
\begin{eqnarray}
D^{-1}_{(0) ab}({\bf q},\omega_{n})= N \left(
 \begin{array}{cccccc}
   \gamma_q & r_{0} & 0 & 0 & 0 & 0 \\
   r_{0} & 0 & 0 & 0 & 0 & 0 \\
   0 & 0 & \frac{4}{J}\Delta^{2} & 0 & 0 & 0 \\
   0 & 0 & 0 & \frac{4}{J}\Delta^{2} & 0 & 0 \\
   0 & 0 & 0 & 0 & \frac{4}{J}\Delta^{2} & 0 \\
   0 & 0 & 0 & 0 & 0 & \frac{4}{J}\Delta^{2} \
 \end{array}
\right)
\end{eqnarray}
\end{widetext}

\noindent
where
$\gamma_{\bf q}=
r_{0}^{2}[ 4V (cos(q_{x})+cos(q_{y}))-4J'cos(q_{x})\;cos(q_{y})]$

The quantities ${\bf q}$ and $\omega_{n}$ are the momentum and the Bose
Matsubara frequency of the boson field, respectively.

We associate with the $N$-component ghost field, ${\bf {\cal Z}}_{p}$, 
the propagator connecting two generic components $p$ and $p'$,

\begin{eqnarray}
B_{pp'} = - \delta_{pp'},
\end{eqnarray}

\noindent
which is $O(1)$

The non-quadratic terms in (A10) define three and four leg
vertices which are $O(1)$.(The full expressions for the vertices
associated with the large-$N$ approach are given in Ref.[\onlinecite{Adriana1}]). In Fig. \ref{fig5} the Feynman rules associated with the large-$N$ approach are summarized.

The charge-charge correlation function is directly associated with
the element $(1,1)$ of the boson propagator $D_{(0)ab} ({\bf
q},\omega_n)$ (the inverse of Eq. (A12)). As in
Ref.[\onlinecite{Jaime4}], up to $O(1/N)$, $D_0(\bf{q},\omega_n)$
is renormalized to $D(\bf{q},\omega_n)$ by an infinite series of
diagrams of $O(1/N)$ (Fig. \ref{fig6}a), and reads:
\begin{equation}
D^{-1}({\bf q},\omega_n)=D^{-1}_0({\bf q},\omega_n)-\Pi({\bf q},
\omega_n),
\end{equation}
where $\Pi({\bf q}, \omega_n)$ is the boson self-energy for which
explicit expressions are given in Ref. \onlinecite{Adriana}.

The superconducting effective interaction between fermions,
$V_{eff}({\bf q},\omega_n)$, can be calculated using the Feynman
rules of Fig. \ref{fig5}. Fig. \ref{fig6}b) represents the diagram
involved in the calculation of $V_{eff}({\bf q},\omega_n)$. The
analytical expression for this diagram is $V_{eff}=\Lambda_a
D_{ab} \Lambda_b$ where $D_{ab}$ is the propagator of the bosonic
field which contains the irreducible self energies of Fig.
\ref{fig6} a and $\Lambda_a$ is the three leg vertex of Fig.
\ref{fig5}. Looking at the order of the propagators and vertices
we see that $V_{eff}({\bf q},\omega_n)$ is $O(1/N)$.

To conclude this appendix we make contact with closely related
approaches such as slave boson formulations.  In contrast to slave
boson theories: (a) Greens functions are calculated in terms of
the original Hubbard operators, (b) fermions, $f_{ip}$, appearing
in the theory are proportional to the Fermi-like $X$-operator
$X^{op}$ (see (A9)) to all orders in the $1/N$
expansion; not only to leading order\cite{Grilli}, (c) as our path
integral is written in terms of $X$-operators we do not need to
introduce {\it a priori} any decoupling scheme, and (d) $r_0$ is
the mean value of $X^{00}$ which is a real field associated with
the number of holes (see Eq. (A8)) and not with the
number of holons. At leading order ($N \rightarrow \infty$ or
$O(1)$) and $V=0$, our formalism is equivalent to slave boson
approaches. However, at the next to leading order ($O(1/N)$),
(which is necessary to calculate one-electron properties such as
the electron self energy $\Sigma({\bf k},\omega)$ and the electron
spectral function $A({\bf k},\omega)$), the two formulations do
not coincide.  The differences between the two formulations are
not yet completely established.  Our theory has the {\it
advantage} that it does not require the introduction
of gauge fields like in slave boson approaches. Hence, through
order $O(1/N)$ we do not need to take care of gauge fluctuations
nor Bose condensation (note that Eq. (A8) does not mean
Bose condensation).  This is important because for the doped
Hubbard model the gauge fluctuations are known to significantly
change the physics.  Careful numerical work will
determine the improvements of the present approach with respect to
slave boson formulations.


\begin{thebibliography}{99}

\bibitem{Ishiguro} T. Ishiguro, K. Yamaji, and G. Saito, {\it
Organic Superconductors}, 2nd ed. (Spinger-Verlag) Berlin (1998).

\bibitem{Jaime1} R. H. McKenzie {\it et. al.}, Phys. Rev. B {\bf 64},
085109 (2001).

\bibitem{Kino} H. Kino and H. Fukuyama, J. Phys. Soc. Jpn. {\bf
65} 2158 (1996).

\bibitem{Seo} H. Seo, J. Phys. Soc. Jpn. {\bf 69}, 805 (2000).

\bibitem{Matteo} M. Calandra, J. Merino and R. H. McKenzie, Phys. Rev. B
{\bf 66 }, 195102 (2002).

\bibitem{Jaime2} J. Merino and R. H. McKenzie, Phys. Rev. Lett. {\bf 87}, 237002 (2001).

\bibitem{Jaime4} J. Merino, A. Greco, R. H. McKenzie, and M. Calandra,
Phys. Rev. B {\bf 68}, 245121 (2003). Note that the captions of
Fig. 15 and Fig. 16 in this reference are interchanged.

\bibitem{Jaime3} M. Dressel {\it et. al.}, Phys. Rev. Lett. {\bf 90}, 167002 (2003).

\bibitem{Ogata} A. Kobayashi, {\it et. al.}, J. Phys. Soc. Jpn. {\bf 73}, 1115 (2004).

\bibitem{Ohta} Y. Ohta, {\it et. al.}, Phys. Rev. B {\bf 50} 13594 (1994).

\bibitem{Adriana} A. Foussats and A. Greco, Phys. Rev. B{\bf 65}, 195107 (2002).

\bibitem{Adriana1} A. Foussats and  A. Greco, Phys. Rev. B (2004), 
{\it in press}.

\bibitem{Grilli} M. Grilli and G. Kotliar, Phys. Rev. Lett. {\bf 64},
1170(1990).

\bibitem{Kotliar} G. Kotliar and J. Liu, Phys. Rev. Lett. {\bf
61}, 1784 (1988).

\bibitem{Vojta} M. Vojta, Phys. Rev. B {\bf 66} 104505 (2002).

\bibitem{Davis} T. Hanaguri, {\it et. al.}, Nature {\bf 430} 1001
(2004). 


\bibitem{Motrunich} O. Motrunich and P. A. Lee, Phys. Rev. B. {\bf 70}, 
024514 (2004).

\bibitem{Andres} A. Greco and R. Zeyher, Eur. Phys. J. B {\bf 6}, 473
(1998).

\bibitem{Coldea} A. I. Coldea, {\it et. al.} submitted to Phys.Rev. B.

\bibitem{Drichko} N. Drichko, {\it private communication}.

\bibitem{Powell} B. J. Powell and Ross H. McKenzie, Phys. Rev. B {\bf 69} 
024519 (2004). 

\bibitem{Hubbard} J. Hubbard, Proc. R. Soc. London, Ser. A{\bf276},238(1963).


\end{thebibliography}
\end{document}